\documentclass[fleqn,usenatbib]{mnras}

\usepackage{amsmath}
\usepackage{newtxtext,newtxmath}
\usepackage{booktabs}
\usepackage{fontawesome}
\usepackage{bm}
\usepackage{fontawesome}
\usepackage{amsbsy}
\usepackage{caption}

\usepackage{soul, CJK}
\usepackage[utf8]{inputenc}

\usepackage{graphicx}

\usepackage{xspace}
\DeclareFixedFont{\ttb}{T1}{txtt}{bx}{n}{9} 
\DeclareFixedFont{\ttm}{T1}{txtt}{m}{n}{9}  
\usepackage{color}
\definecolor{deepblue}{rgb}{0,0,0.5}
\definecolor{deepred}{rgb}{0.6,0,0}
\definecolor{deepgreen}{rgb}{0,0.5,0}
\usepackage{listings}
\usepackage{gensymb}

\newcommand\im{\mathrm i}
\newcommand\pythonstyle{\lstset{
language=Python,
basicstyle=\ttm,
otherkeywords={self},             
keywordstyle=\ttb\color{deepblue},
emph={MyClass,__init__},          
emphstyle=\ttb\color{deepred},    
stringstyle=\color{deepgreen},
frame=tb,                         
showstringspaces=false            %
}}
\lstnewenvironment{pyt}[1][]
{
\pythonstyle
\lstset{#1}
}
{}

\newcommand{\zetaa}{\zeta^{[2]}}
\newcommand{\zz}{z^{[2]}}

\newcommand{\br}[1]{\mathopen{}\left(#1\right)\mathclose{}}

\newcommand{\bigo}[1]{\mathcal{O}\br{#1}}

\usepackage{listings}

\graphicspath{{./}{figures/}}

\title[Perturbative Picture of Planetary Microlensing]{On the Perturbative Picture and the Chang-Refsdal Lens Approximation for Planetary Microlensing}

\author[K. Zhang]{
Keming Zhang \begin{CJK*}{UTF8}{gkai}(张可名)\end{CJK*}$^{1}$\thanks{E-mail: kemingz@berkeley.edu}
\\
$^{1}$Department of Astronomy, University of California, Berkeley, CA 94720-3411, USA\\
}

\date{Accepted XXX. Received YYY; in original form ZZZ}

\begin{document}

\label{firstpage}
\pagerange{\pageref{firstpage}--\pageref{lastpage}}
\maketitle

\begin{abstract}

Under the perturbative picture of planetary microlensing, the planet is considered to act as a uniform-shear Chang-Refsdal lens on one of the two images produced by the host star that comes close to the angular Einstein radius of the planet, leaving the other image unaffected. However, this uniform-shear approximation is only valid for isolated planetary caustics and breaks down in the resonant regime.
Recently, the planetary-caustic degeneracy arising from the above formalism is found to generalize to the regime of central and resonant caustics, indicating that the perturbative picture and Chang-Refsdal lens approximation may have been under-explored in the past.
Here, I introduce a new variable-shear Chang-Refsdal lens approximation, which not only supports central and resonant caustics, but also enables full magnification maps to be calculated analytically.
Moreover, I introduce the generalized perturbative picture, which relaxes the required proximity between the planet and the image being perturbed in the previous work.
Specifically, the planet always perturbs the image in the same half of the lens plane as the planet itself, leaving the other image largely unaffected.
It is demonstrated how this new framework results in the offset degeneracy as a consequence of physical symmetry.
The generalized perturbative picture also points to an approach to solve the two-body lens equation semi-analytically.
The analytic and semi-analytic microlensing solutions associated with this work may allow for substantially faster light-curve calculations and modeling of observed events.
A python implementation is provided.
\end{abstract}


\section{Introduction}
\label{sec:intro}
In the simplest microlensing scenario, a foreground lens star splits a background source star into two images that are located inside and outside the Einstein radius of the lens star,
\begin{equation}
\label{eq:eins}
    \theta_{\rm E}=\sqrt{\dfrac{4GM}{D_{\rm rel}c^2}},
\end{equation}
where $G$ is the gravitational constant, $M$ is the lens mass, $c$ is the speed of light, and $D_{\rm rel}^{-1}=D_{\rm lens}^{-1}-D_{\rm source}^{-1}$ is related to the relative distance between the lens and source. The image outside the Einstein ring is usually 
referred to as the major image and the inside image as the minor image. The locations of the major/minor images, along with their magnifications, can also be expressed as simple closed-form expressions of the source location.

A two-body lens, on the other hand, splits a source star into either three or five images, depending on whether the source is inside or outside of caustics. The locations of the images are found by solving the lens equation in its complex form \citep{witt_investigation_1990}
\begin{equation}
    \zeta = z-\dfrac{1}{\bar{z}}-\dfrac{q}{\bar{z}-s},
    \label{eq:lens1}
\end{equation}
where $\zeta=\xi+\im \eta$ is the true source location, $z=z_1+\im z_2$ is the image location, $q$ is the mass ratio between the two lens components, and $s$ their projected separation in units of the Einstein ring radius of the more massive lens component. The above equation can be transformed into a quintic polynomial that can only be solved numerically. As pointed out in \cite{witt_minimum_1995}, the fact that the binary lens equation is not analytically tractable presents a major obstacle in further analytical studies. Additionally, when finite source effects are considered, this quintic polynomial generally has to be solved repeatedly to account for the variance of magnification over the source area, thereby creating a computationally non-trivial problem for the modeling of observed events.

The resemblance between the two-body lens with planetary mass ratios ($q\ll1$) and the Chang-Refsdal lens has provided one pathway toward analytic studies of planetary microlensing. Both the planetary lens and the Chang-Refsdal lens consist of two components with extreme mass ratios. The Chang-Refsdal lens describes a point-mass lens perturbed by uniform external shear, and was introduced by \cite{chang_flux_1979} to describe the action of an individual star on the outskirts of a massive galaxy acting as a gravitational lens on a background quasar. 
For a Chang-Refsdal lens, a star lying close to a given quasar image could produce a time-variable magnification to that image due to the relative proper motion between the star/galaxy and quasar. This has led to the ``perturbative picture'' of planetary microlensing \citep{gould_discovering_1992,gaudi_planet_1997}, where a planetary-mass body acts as a uniform-shear Chang-Refsdal lens on one of the major/minor images produced by the primary star that comes close to it of order its angular Einstein radius, leaving the other image unaffected.

One advantage of the Chang-Refsdal lens approximation is that the Chang-Refsdal lens equation can be transformed into a quartic polynomial, which is the highest-order polynomial that can be solved analytically. The Chang-Refsdal lens equation in  its complex form is written as
\begin{equation}
\label{eq:pure-shear}
    \zeta = z-\dfrac{1}{\bar{z}}+\gamma \bar{z},
\end{equation}
where $\gamma$ denotes the shear. This property has allowed for extensive analytic studies of the Chang-Refsdal lens, notably in \cite{an_changrefsdal_2006}. 

However, there are important differences between the planetary lens and the Chang-Refsdal lens, which substantially limit the validity of the approximation of the former by the latter. While the star-galaxy mass ratio for the Chang-Refsdal lens is often below $q=10^{-12}$, the mass ratio for planetary lenses could range anywhere between $q\sim10^{-2}$ for Jovian planets and $q\sim10^{-5}$ for terrestrial planets. 
The $\sqrt{M}$ scaling (Equation \ref{eq:eins}) of the Einstein ring radius suggests the Einstein radius of the planet may be a substantial fraction of that of the primary star.
Thus, the effects of the primary star often can hardly be considered as a uniform background shear over the sphere of influence of the planet.
Indeed, \cite{dominik_binary_1999} pointed out that the Chang-Refsdal approximation is only valid for planets sufficiently far from the Einstein ring of the primary star, where the effect of the planetary caustics can be considered in isolation and as a Chang-Refsdal caustic.

Moreover, the appreciable mass ratio of the planetary microlens has allowed for the existence of central and resonant caustics, which are not allowed under the Chang-Refsdal lens formalism. Additionally, planetary caustics in practice are usually elongated towards the host star along the real axis, whereas the Chang-Refsdal caustic is completely symmetrical.
There exist other analytical studies that take advantage of the planetary mass ratio ($q\ll1$), which has led to interesting results (e.g.\ \citealt{bozza_perturbative_1999,bozza_caustics_2000,an_gravitational_2005}). Nevertheless, to date, there has been an absence of an analytic framework for planetary microlensing that holds for all types of caustic topologies. As a result, modeling of current observations still relies on numerically solving the full lens equation, for which optimized quintic solvers have been developed that provide order-unity speed up \citep{skowron_general_2012,fatheddin_improved_2022} compared to a baseline \texttt{ZROOTs} routine from \textit{Numerical Recipes}.

Recent results in microlensing degeneracy indicate that the Chang-Refsdal approximation and the perturbative picture may have been under-explored in the past.
Specifically, an important consequence of the Chang-Refsdal approximation is the existence of light-curve degeneracies for planetary caustic perturbations \citep{gaudi_planet_1997}, commonly referred to as the inner-outer degeneracy \citep{han_moa-2016-blg-319lb_2018}.
Here, the source trajectory is expected to pass equidistant to the planetary caustics (located at $s-1/s$) of the degenerate lens configurations, owing to its symmetry under the Chang-Refsdal lens approximation.
However, observed degeneracies that reference the inner-outer degeneracy rarely have well-isolated planetary caustics \citep{yee_ogle-2019-blg-0960_2021}, although the degenerate light curves often have excellent resemblance.
Recently, the offset degeneracy proposed by \cite{zhang_ubiquitous_2022}  found the equidistance relationship underlying the inner-outer degeneracy to also apply to bi-modal solutions usually attributed to the close-wide degeneracy for central caustics, along with degeneracies involving two resonant topology solutions (cf.\ \citealt{gould_systematic_2022}).

The condition that the source trajectory shall pass equidistant to the locations $s-1/s$ of the degenerate solutions regardless of the caustic topology was recently proved in \cite{zhang_mathematical_2022}.
However, the interpretation of the $s-1/s$ term as the location of the planetary caustic is rather unsatisfactory, especially for light-curve anomalies primarily associated with central and resonant caustics.
Under the perturbative picture, the $s-1/s$ term describes the coordinate origin of the Chang-Refsdal lens approximation.
The fact that the equidistance relationship with respect to $s-1/s$ persists for central and resonant caustics therefore suggests the possibility that the perturbative picture and the Chang-Refsdal lens approximation may also be generalized beyond the planetary caustic, which is studied in the current work.

This paper is organized as follows. In Section \ref{sec:perturb}, I introduce the generalized perturbative picture, which relaxes the required proximity between the planet and the image being perturbed in the previous work. Specifically, the planet always perturbs the image in the same half of the lens plane as the planet itself, leaving the other image largely unaffected.
In Section \ref{sec:cr}, I propose a new variable-shear Chang-Refsdal lens approximation that quantifies the generalized perturbative picture. I show that the offset degeneracy becomes a consequence of physical symmetry under this variable-shear approximation. Crucially, the proposed Chang-Refsdal formalism enables full magnification maps to be derived analytically, whose accuracy is examined in Section \ref{sec:mag}. Section \ref{sec:caus} shows that this variable-shear Chang-Refsdal lens formalism recovers known caustic properties of the planetary lens. In Section \ref{sec:semi}, I introduce a semi-analytic approach to solve the lens-equation exactly that is associated with the generalized perturbative picture. The results of this paper are reviewed in Section \ref{sec:disc}, where I discuss how they may be employed to substantially accelerate the modeling of observed events. A \texttt{Python} implementation of the exact semi-analytic and approximate analytic microlensing solutions presented in Sections \ref{sec:mag} \& \ref{sec:semi} is provided\footnote{https://github.com/kmzzhang/analytic-lensing}.

\section{The Perturbative Picture}
\label{sec:perturb}
\begin{figure*}
 \centering
 \includegraphics[width=\textwidth]{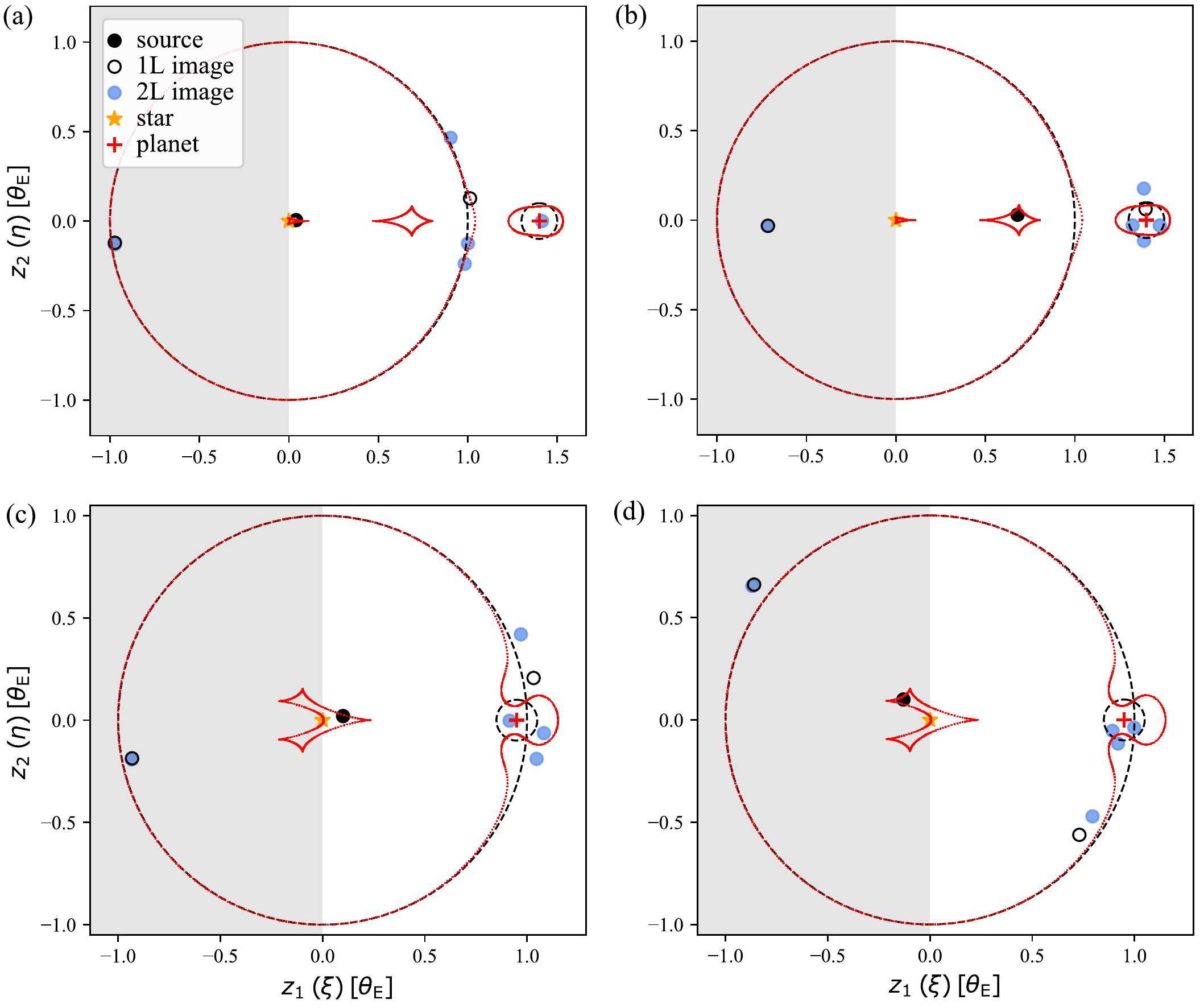}
 \caption[Illustration of the generalized perturbative picture for central, planetary, and resonant caustic perturbations]{Illustration of the generalized perturbative picture for (a) central, (b) planetary, and (c,d) resonant caustic perturbations. (a,b,c) show major image perturbations and (d) shows minor image perturbation. The sphere of influence of the planet in the lens plane is indicated by the non-shaded region to the right with $z_1>0$. In the legend, 1L refers to the single-lens major/minor images resulting from the primary star alone, and 2L refers to the five images produced by the star-planet binary lens. In each case, the image in the shaded region is shown to be largely unaffected by the presence of the planet, as the 1L and 2L images coincide. The critical curves and caustics are shown in red, and the angular Einstein ring radius for the star and planet are shown in black dashed lines for comparison. The source star (black dot) is inside the caustics for all three cases. The mass ratio is $q=0.01$ and the projected separation is $s=1.4$ for (a,b) and $s=0.95$ for (c,d).}
 \label{fig:perturb}
\end{figure*}

\begin{figure*}
 \centering
 \includegraphics[width=\textwidth]{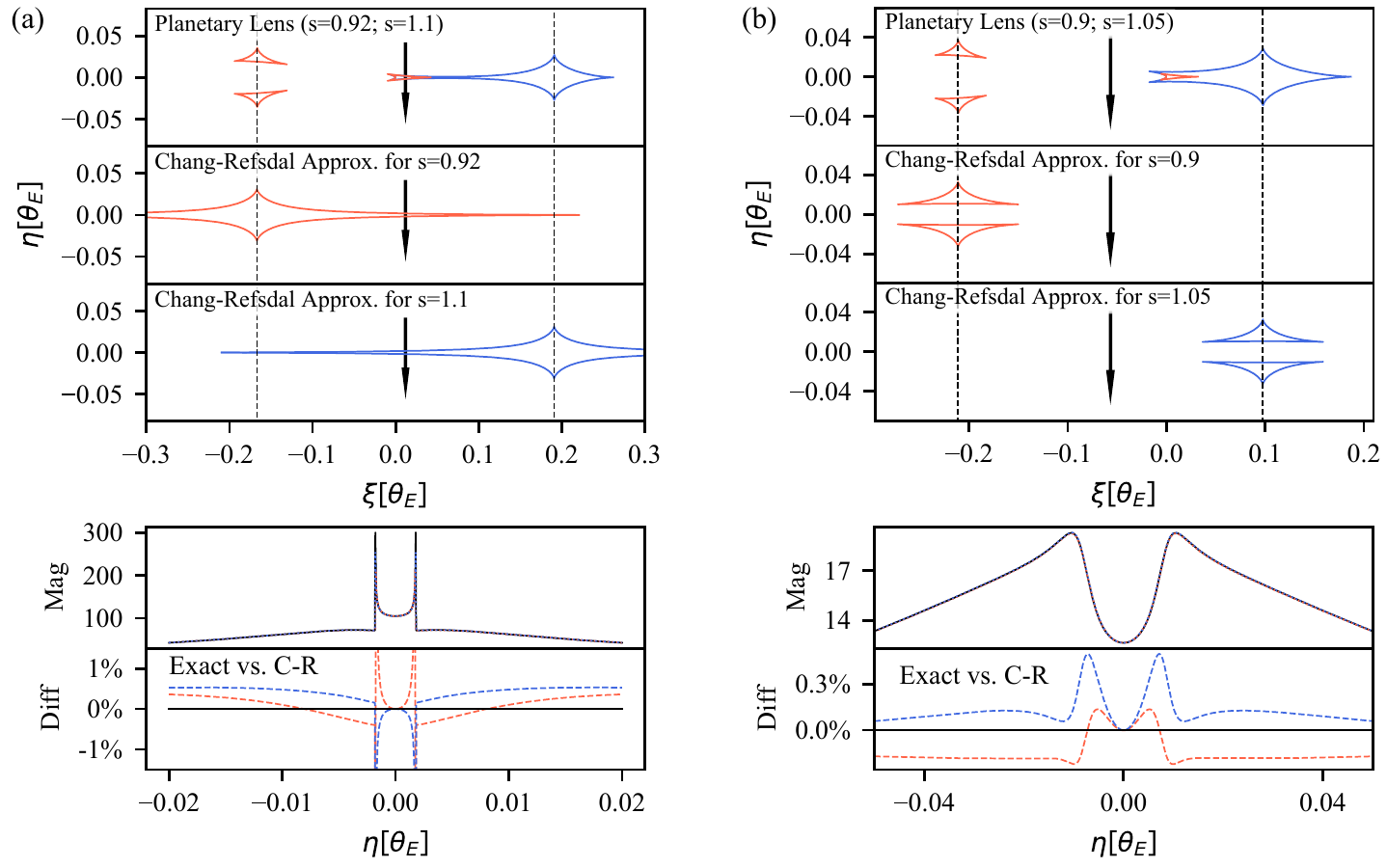}
 \caption[Illustration of the Chang-Refsdal lens approximation in the context of the Offset Degeneracy]{Illustration of the Chang-Refsdal lens approximation in the context of the Offset Degeneracy for (a) the generalized major image perturbation and (b) the generalized minor image perturbation. The top panel of (a,b) shows the caustics of two lens configurations overlaid in red ($s<1$; close) and blue ($s>1$; wide), which give rise to degenerate light curves (second-from-bottom panel) for the source trajectory (vertical arrow) that crosses equidistance to the locations $s-1/s$ (dashed lines) of the two lens configurations. The 2nd and 3rd panels from the top illustrate the Chang-Refsdal lens approximation for each lens configuration, which are shown to be exactly symmetrical with respect to the source trajectory. The bottom panels show the differences from the approximate Chang-Refsdal light curve to the two exact point-source light curves, shown in the same color coding. The mass ratio is $q=5\times10^{-4}$ for both subplots.}
 \label{fig:illus1}
\end{figure*}

The perturbative picture for planetary microlensing was initially laid out in \cite{gould_discovering_1992}, which states ``[a] planet of mass $m$ affects appreciably the microlensing image only if the planet and the unperturbed image are separated by or order the planet's own Einstein radius,'' and that ``at most one image is significantly affected and that the perturbed images lie near the unperturbed image[.]''
Subsequent work then showed that the planet could also appreciably affect the microlensing image even if the planet lies far from the unperturbed image\footnote{In this work, the term perturbation is used solely with respect to the planet. Thus, the unperturbed image refers to the major/minor images resulting from the primary star alone.}, namely via central caustics for high magnification events \citep{griest_use_1998}.
The question that remains is whether the condition ``at most one image is significantly affected'' holds when the source passes close to central and resonant caustics.

To answer this question, it is illuminating to consider the lens plane of microlensing as opposed to the source plane. The lens plane describes the perturbing lens masses, the resulting images, and the critical curves, whereas the source plane describes magnification maps along with caustics resulting from all contributing images. Thus, the lens plane describes the \textit{cause} and the source plane describes the \textit{effect}. In the source plane, the proximity between the source and caustics correlates with higher magnifications. Analogously, the proximity between images and critical curves serves a similar purpose. For example, for a single lens, there exists one point-like caustic exactly at the lens mass itself. As the source approaches this singularity, both major/minor images approach the critical curve, leading to increased magnification. When the source coincides with the caustic, the image also coincides with the critical curve as an Einstein ring, which results in infinite magnification for a point source.

Let us now consider how the presence of the planet as an additional lens mass causes the critical curve to deviate from the Einstein ring of the primary mass.
As illustrated in Figure \ref{fig:perturb}(a, b), in the non-resonant regime, there is an isolated ``planetary'' critical curve centered on the planet with spatial scale $\theta_{\rm E,p}$. Parts of the ``primary'' critical curve ($\theta_{\rm E,\star}$) near its intersection with the positive real axis are elongated towards the planet, which is associated with the existence of central caustics. 
In the resonant regime (Figure \ref{fig:perturb}c,d), the primary and planetary critical curves merge.

From these examples, it can be seen that the planet only affects parts of the critical curve in the positive lens plane, leaving the negative lens plane (shaded regions in Figure \ref{fig:perturb}) largely unaffected.
By considering the proximity of the unperturbed image locations to the critical curves as a proxy for magnification, we may then conclude that the planet also only perturbs the single-lens image in the positive lens plane, leaving the image in the negative lens plane unaffected. As illustrated in Figure \ref{fig:perturb}, this is indeed the case regardless of the proximity between the planet and the image being perturbed.

While the original perturbative picture considers the sphere of influence of the planet as limited to its angular Einstein radius, the above discussion indicates a generalized perturbative picture: the sphere of influence of the planet is constrained to one-half of the lens plane, where it splits one of the major/minor images into two or four images, leaving the other image largely unaffected.
This generalized perturbative picture then allows for a unified classification of planetary perturbations into major-image perturbations and minor-image perturbations.
Under the original perturbative picture, the distinction between major and minor-image perturbations has been restricted to planetary-caustic perturbations (Figure \ref{fig:perturb}b). Thus, major-image perturbations have been restricted to wide-separation planets ($s>1$) and vice versa\footnote{See the Appendix in \cite{han_moa-2016-blg-319lb_2018}: ``Types of Planetary Perturbations''}.
The present discussion shows that the distinction between major/minor image perturbations should be made not by the location of the planet, but by the location of the source instead.
Specifically, major-image perturbations occur when the source is in the positive source plane ($\zeta>0$), and vice versa. This would then allow major-image perturbations to be generalized to $s<1$ planets, as illustrated in Figure \ref{fig:perturb}(c).

\section{The Chang-Refsdal Lens Approximation}
\label{sec:cr}

The uniform-shear Chang-Refsdal lens approximation quantifies the action of the planet under the original perturbative picture of \cite{gould_discovering_1992}.
Here, I introduce a variable-shear Chang-Refsdal lens approximation that quantifies the generalized perturbative picture, which accurately describes all of the central, resonant and planetary caustics.

\subsection{Uniform-Shear Approximation}

Since past works have adapted slightly different conventions on the uniform-shear Chang-Refsdal lens approximation, let us first re-examine the relevant works using a uniform notation of the complex two-body lens equation (Equation \ref{eq:lens1}). To consider the lensing behavior near the planetary lens companion, let us first transform the complex lens equation from the \textit{primary} frame ($\zeta$, $z$) to the \textit{planetary} frame ($\zetaa$, $\zz$), which has units of the planetary Einstein radius $\theta_{\rm E,p}=\sqrt{q}\theta_{\rm E,\star}$, and coordinate origins at the location of the planet ($z=s$) for the lens plane, with the corresponding location in the source plane ($\zeta=s-1/s$). The latter is often interpreted as the location of the planetary caustic. Applying the coordinate transformation
\begin{align}
\label{eq:trans}
    \zeta &= \sqrt{q}\zeta^{[2]}+s-1/s\nonumber\\
    z &= \sqrt{q}\zz+s,
\end{align}
and rearranging, the two-body lens equation becomes
\begin{equation}
\label{eq:tr}
    \zeta^{[2]}=\zz-\dfrac{1}{\bar{z}^{[2]}}+\dfrac{\bar{z}^{[2]}}{s\cdot(\sqrt{q}\bar{z}^{[2]}+s)}.
\end{equation}

In the limit of $q\rightarrow0$, the above equation is reduced to the Chang-Refsdal lens with uniform shear $\gamma=1/s^2$ \citep{dominik_binary_1999}. For finite $q\ll1$, the Chang-Refsdal lens with $\gamma=1/s^2$ is the first order Taylor expansion of Equation \ref{eq:tr} around $\bar{z}^{[2]}=0$ \citep{dominik_binary_1999, bozza_caustics_2000}, which can also be interpreted as a power-series in $\sqrt{q}$
\begin{equation}
\label{eq:exp}
     \zeta^{[2]}=\zz-\dfrac{1}{\bar{z}^{[2]}}+\sum_{i=1}^{\infty}(-1)^{i+1}\cdot q^{(i-1)/2}\cdot\dfrac{(\bar{z}^{[2]})^i}{s^{i+1}}.
\end{equation}

On the other hand, the original Chang-Refsdal approximation of the earlier work of \cite{gaudi_planet_1997} adopted a slightly different shear definition. 
Instead of the planet location, the shear is evaluated at the location of the image being perturbed at the mid-point of the perturbation, which occurs when the source crosses the star-planet axis.
Recall that the original perturbative picture requires the image being perturbed to pass the planet closer than $\bigo{\theta_{E,\rm p}}$.
Therefore, the location of the image being perturbed would approach the planet location for $q\rightarrow0$, and the two shear definitions would become equivalent.

\subsection{Variable-Shear Approximation}
\label{sec:lens}

In this subsection, I introduce a new variable-shear Chang-Refsdal lens approximation that holds for all three caustic topologies.
The shear is defined to be real positive $\gamma=1/z_+^2$, where
\begin{equation}
\label{eq:shear}
    z_+=\dfrac{\sqrt{\xi^2+4}+\xi}{2}.
\end{equation}

For sources on the real axis ($\zeta=\xi$), the shear definition is identical to \cite{gaudi_planet_1997}, which is nevertheless formulated within the original perturbative picture that requires $|z_+-s|\lesssim\bigo{\theta_{E,\rm p}}$. Under the generalized perturbative picture, Equation \ref{eq:shear} corresponds to the unperturbed location of the image that is assumed to be perturbed by the planet, which is the major image for $\xi>0$ and the minor image for $\xi<0$.
The image being perturbed is always in the positive lens plane, and thus the ``+'' subscript.
For sources off the real axis ($\eta\neq0$), the shear is evaluated by projecting the source location onto the real axis. Thus the lines of constant shear (LCS) are perpendicular to the real axis by construction. 

The proposed approximation is different from the variable-shear approximation of \cite{gould_discovering_1992}, which evaluates the shear directly at unperturbed image location rather than its projection on the real axis. The formalism of this earlier work was derived by Taylor expanding the time-delay surface at the unperturbed image location, which was motivated by the condition where ``the perturbed images lie near the unperturbed image[.]'' Again, this assumption only holds for isolated planetary caustics, as can be seen in Figure \ref{fig:perturb}. As noted in footnote 3 of \cite{gould_discovering_1992}, this approximation also results in a leftward arching of the planetary caustics that is not present in the exact calculation, nor the new variable-shear formalism proposed here (see Section \ref{sec:mag}).

In contrast, the requirement of the shear to be real and the LCS to be perpendicular to the star-planet axis is motivated by conditions of the offset degeneracy. 
As illustrated in Figure \ref{fig:illus1}, vertical source trajectories result in nearly identical light curves under the degenerate lens configurations \citep{zhang_ubiquitous_2022}. Now, if one were to apply a literal reading of the uniform-shear approximation of \cite{gaudi_planet_1997} and hold the shear fixed on the star-planet axis, one would find that the resulting light curve under the Chang-Refsdal lens approximation nearly perfectly resembles both of the degenerate light curves. In fact, the extent to which the Chang-Refsdal light curves deviate from the exact light curves is similar to the extent to which the two degenerate light curves deviate from one another.

The above findings may appear rather surprising, since the uniform-shear approximation of \cite{gaudi_planet_1997} is known to fail for resonant and central caustics \citep{dominik_binary_1999}. However, an implicit assumption made in the previous works is that the uniform-shear approximation fails as a \textit{global} approximation. As we have seen from Figure \ref{fig:illus1}, the uniform-shear Chang-Refsdal lens serves as an excellent \textit{local} approximation along the vertical direction in the source plane, despite the absence of \textit{global} caustic resemblance. Consequently, for oblique trajectories, one can derive equally accurate light-curve approximations by evaluating the shear at the projection of the source onto the lens axis. 

As I will show, this variable-shear Chang-Refsdal lens approximation not only leads to accurate magnification maps (Section \ref{sec:mag}), but also recovers known caustic properties of the planetary lens (Section \ref{sec:caus}). However, the accuracy of the proposed variable-shear approximation in the source plane contrasts sharply with the fact that it does not recover the correct image locations. This can be easily seen by considering a hypothetical source at infinity, where the major image overlaps with the true source location. Here, the shear goes to zero and the Chang-Refsdal lens is reduced to the point lens, but the origins of the primary and planetary frames differ by $1/s$, thus placing the major image at the wrong location. This behavior indicates that the variable-shear lens should be considered degenerate with the exact planetary lens because they share similar source-plane but not lens-plane behavior. This intriguing behavior deserves further analytical study in future works.

\section{Analytic Magnifications}
\label{sec:mag}

\begin{figure*}
 \centering
 \includegraphics[width=\textwidth]{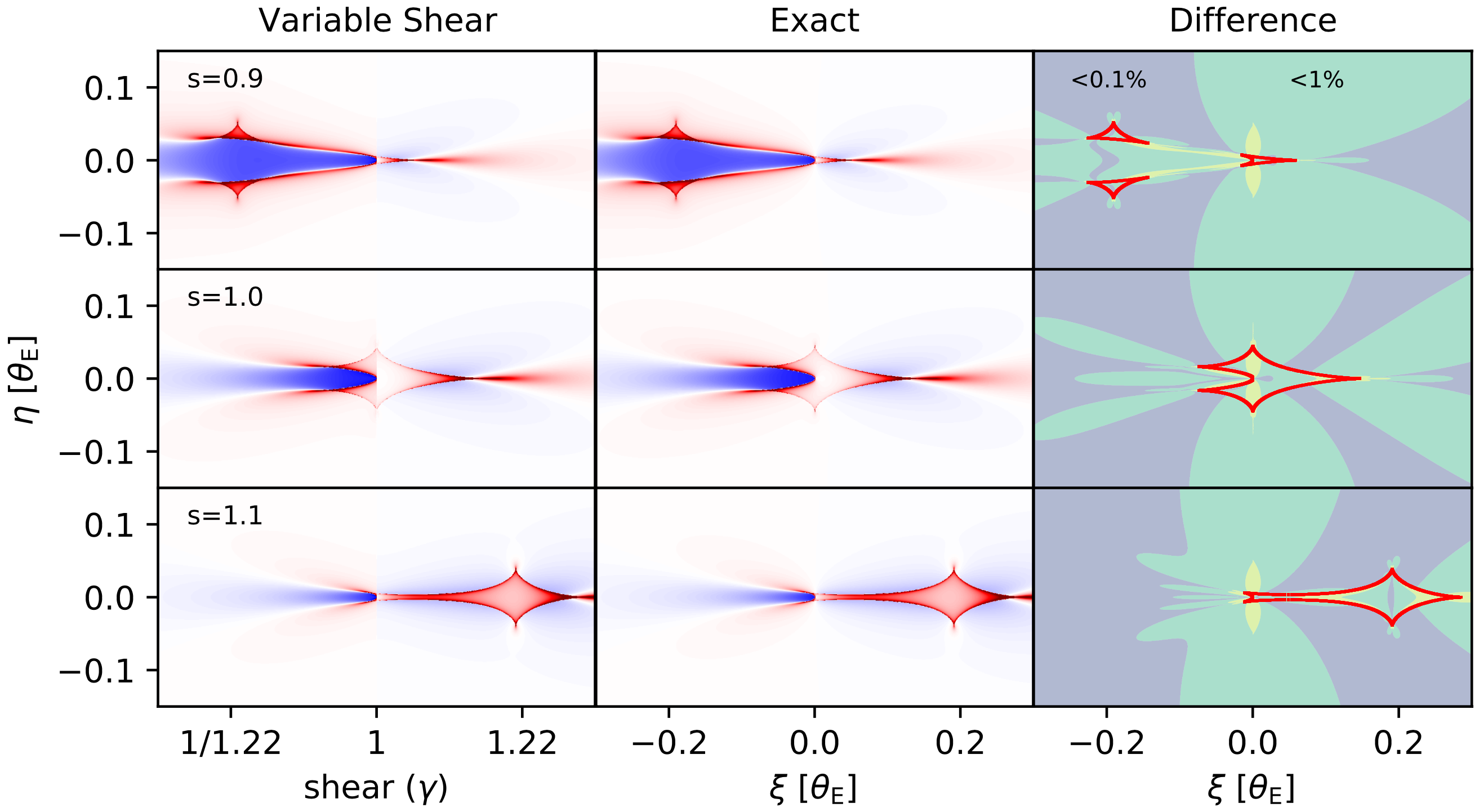}
 \caption[Variable-shear and exact magnification maps in the resonant regime.]{Magnification maps calculated using the variable-shear Chang-Refsdal lens approximation and the exact lens formalism (with the semi-analytic solver of Section \ref{sec:semi}), as well as their fractional differences. The magnification maps are visualized as the deviation from the point-lens point-source magnification map, where excess magnification is shown in red and suppressed magnification is shown in blue, with consistent color scale across subplots. The horizontal axis for variable shear is re-parameterized with shear ($\gamma$) using the definition in Equation \ref{eq:shear}. The coordinate origin for the exact calculation is offset to the primary cusp. The color coding in the difference maps indicates differences of less than 0.1\%, 1\%, 10\%, as labeled in the middle panel except for the light-green central region of <10\%. The caustics are overlaid in red in the difference maps for reference. The mass ratio is $q=10^{-3}$ for all subplots.}
 \label{fig:mag-resonant}
\end{figure*}

Given that the Chang-Refsdal lens equation can be transformed into a quartic polynomial, we may now calculate full magnification maps for the planetary lens analytically. To acquire analytic magnifications, one first takes the complex conjugate of Equation \ref{eq:pure-shear}, and substitutes the expression for $\bar{z}$ back into Equation \ref{eq:pure-shear} itself. After clearing fractions, we arrive at a quartic polynomial,
\begin{equation}
    p(\zz)=\sum_{i=0}^{4} a_i(\zetaa,\bar{\zeta}^{[2]},\gamma)\cdot (\zz)^i=0,
    \label{eq:quartic}
\end{equation}
where, with $\gamma=1/z_+^2$ (Equation \ref{eq:shear}),
\begin{align}
    a_0=& \gamma\nonumber\\
    a_1=&-\zetaa + 2\gamma \bar{\zeta}^{[2]}\nonumber\\
    a_2=& -2\gamma^2 - \zetaa\bar{\zeta}^{[2]} + \gamma \bar{\zeta}^{[2]^2}\nonumber\\
    a_3=&\gamma\zetaa + \bar{\zeta}^{[2]} - 2\gamma^2 \bar{\zeta}^{[2]}\nonumber\\
    a_4=&-\gamma + \gamma^3\nonumber.
\end{align}

Note that not all roots of the quartic polynomial are solutions to the original lens equation, and each solution should be verified by plugging back into Equation \ref{eq:pure-shear}. The total magnification is the sum of the magnification of each individual image, which is given by the absolute value of the inverse Jacobian determinant,
\begin{equation}
    \mu_\gamma=\sum_j\left|1-\dfrac{\partial \zetaa}{\partial \bar{z}^{[2]}}\dfrac{\overline{\partial \zeta}^{[2]}}{\partial \bar{z}^{[2]}}\right|^{-1}_{z^{[2]}_j},
\end{equation}
where the derivatives are evaluated using Equation \ref{eq:pure-shear} at the valid image solutions $z^{[2]}_j$. The variable-shear lens provides the magnification perturbation by the planet through,
\begin{align}
    \Delta\mu&=\mu_\gamma - \mu_\infty\\
    &=\mu_\gamma - \dfrac{1}{|\gamma^2-1|},
\end{align}
where $\mu_\infty$ is the terminal magnification ($|\zetaa|\rightarrow\infty$). With $u=|\zeta|$, the full magnification for the total of three or five images can be found by adding back the single-lens magnifications,
\begin{equation}
\label{eq:mag}
    \mu=\dfrac{u^2+2}{u\sqrt{u^2+4}}+\Delta\mu.
\end{equation}

Although lengthy when expressed as a function of ($\zeta, s, q$), Equation \ref{eq:mag} is indeed closed-form and may offer substantial speed-up in the calculation of planetary microlensing light curves and the modeling of observed events. Let us now examine the accuracy of magnification maps under the variable-shear Chang-Refsdal approximation.
It is already known from previous works that the Chang-Refsdal lens provides excellent approximation near isolated planetary caustics. Therefore, let us first examine the accuracy of the variable-shear magnification maps near resonant and semi-resonant caustics, before examining magnification maps near central caustics.

\begin{figure*}
 \centering
 \includegraphics[width=\textwidth]{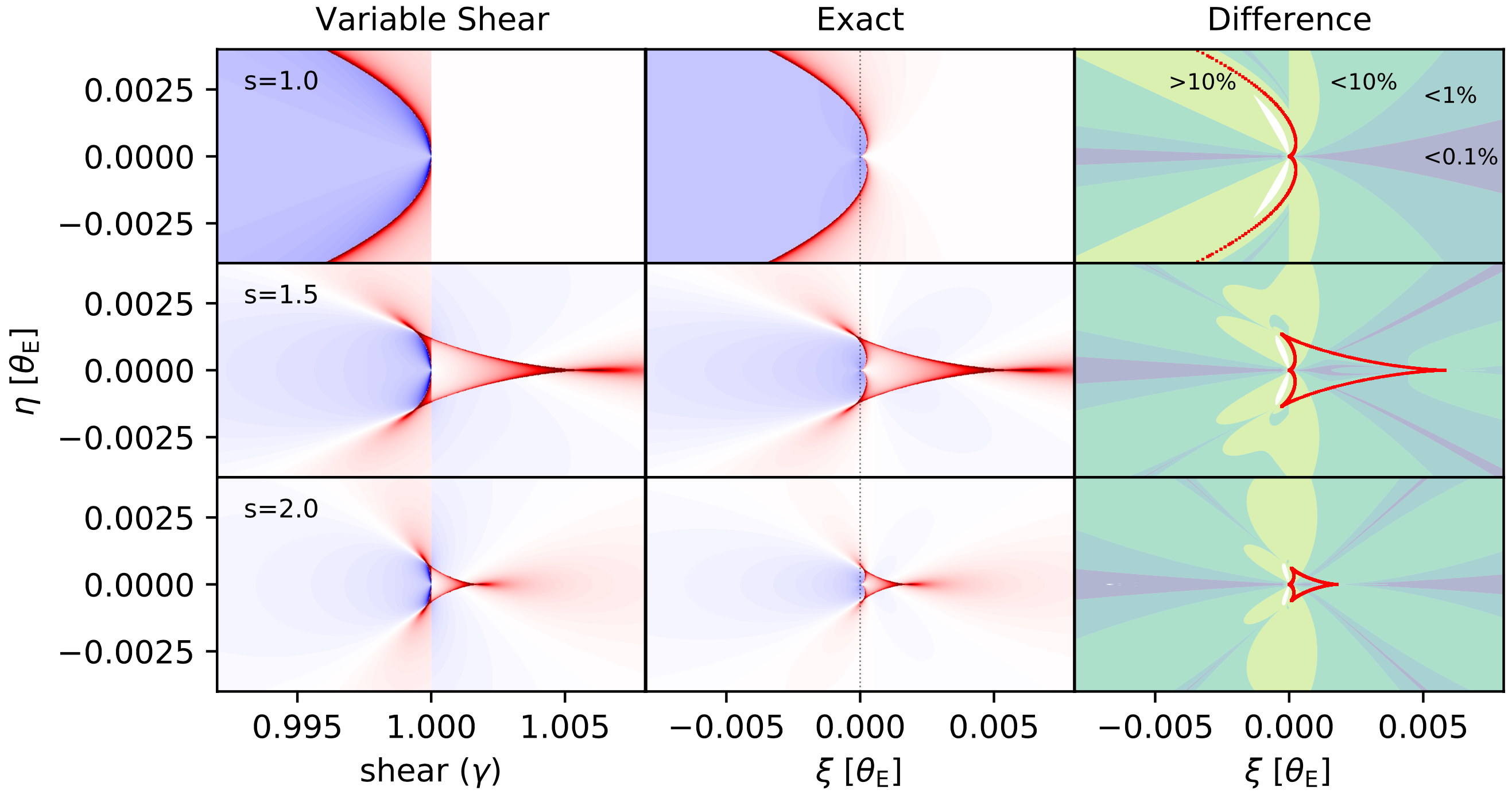}
 \caption[Variable-shear and exact magnification maps in the high-magnification regime.]{Similar to Figure \ref{fig:mag-resonant}, but zoomed into the central region and shown for $s=(1,1.5,2)$. The color scale is consistent across subplots, but different from Figure \ref{fig:mag-resonant}. The coordinate origin for the exact calculation is offset to the primary cusp. The uncolored regions in the difference maps indicate negative magnifications resulting from the variable-shear approximation.}
 \label{fig:mag-central}
\end{figure*}

Figure \ref{fig:mag-resonant} shows the variable-shear and the exact calculations of magnification maps for lenses in or near the resonant regime, which appear nearly identical. The magnification difference maps reveal two major regimes where the two calculations differ by $>1\%$. First, there is a dumbbell-shaped structure along the imaginary axis of size $\Delta\eta\sim0.1$, the interpretation of which will be clear with the discussion of Figure \ref{fig:mag-central} in the next paragraph.
Deviations greater than $1\%$ also occur along the excess magnification ridges straddling the suppressed magnification zone between the close-planetary caustic and the central caustic, as seen in the top panel of Figure \ref{fig:mag-resonant}. This type of deviation concerns the exact shape of the excess magnification ridges. One example is already seen in Figure \ref{fig:illus1}(b) where, with a mass ratio of $q=5\times10^{-4}$, the maximum deviation is merely $0.5\%$.
Therefore, the variable-shear approximation is in excellent agreement with the exact calculation outside of a small central region, which we turn our attention to now.

\begin{figure*}
 \centering
 \includegraphics[width=\textwidth]{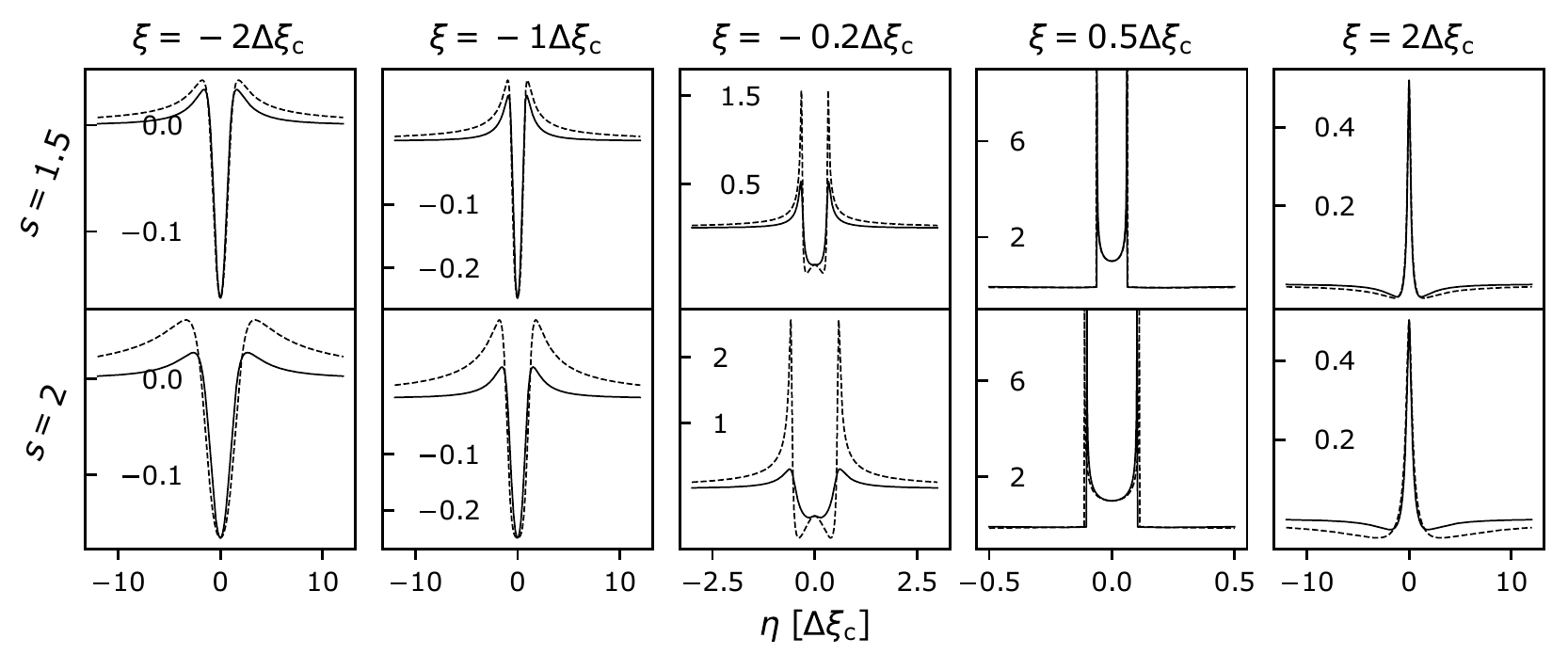}
 \caption[Magnification slices (light curves) along the vertical direction for $s=(1.5, 2)$]{Magnification slices (light curves) along the vertical direction for $s=(1.5, 2)$, which are associated with the lower two panels of Figure \ref{fig:mag-central}. As shown in the subplot titles, the impact parameters of the magnification slices are in units of the central-caustic size, which is $\Delta\xi_{\rm c}\sim0.006$ for $s=1.5$ and $\Delta\xi_{\rm c}\sim0.002$ for $s=2$. The exact calculation is shown in solid lines and the variable-shear calculation in dashed lines. The vertical axes are in units of the fractional deviation from the point-source point-lens magnification.}
 \label{fig:lc-central}
\end{figure*}

Both of the aforementioned types of discrepancies become more pronounced in the high-magnification regime. In Figure \ref{fig:mag-central}, one immediately notices the discontinuity across the imaginary axis for the variable-shear calculation, which accounts for the dumbbell-shaped structure seen in Figure \ref{fig:mag-resonant}. For sources near the imaginary axis and the primary star, the two unperturbed image locations are about equidistant to the planet and both images are substantially affected by the planet. In other words, the generalized perturbative picture is no longer accurate in the high-magnification regime near the imaginary axis. The magnification near the imaginary axis is overestimated for minor-image perturbations and underestimated for major-image perturbations. 
Given that this discontinuity is known in advance and does not usually coincide with true planetary features, one may apply \textit{post-hoc} corrections when applying the analytic magnification to the modeling of observed events, for example, by reducing the weights of photometric data points near the imaginary axis.

The discrepancy along the excess magnification ridge becomes prominent for high-magnification minor-image perturbations, where high magnification means the immediate vicinity of central caustics with impact parameters of $u_0\sim q$. Figure \ref{fig:lc-central} shows magnification slices along the vertical direction with various impact parameters in units of the central-caustic size (Equation \ref{eq:xic}).
The left two columns show that the variable-shear calculation substantially overestimates the strength of the excess magnification ridge.
The middle panel of Figure \ref{fig:lc-central} shows that this deviation diverges in the immediate vicinity of the two off-axis central-caustic cusps, where minuscule differences in the caustic shape becomes important.
The magnification difference maps in Figure \ref{fig:mag-resonant} also reveal limited regions of spatial scale $\bigo{q}$ with unphysical negative magnifications under the two caustic folds towards the back end, which is also reflected in the double dips in the middle panels of Figure \ref{fig:lc-central}. The interpretation of these behaviors will become clearer with the discussion of caustics in Section \ref{sec:caus}.

\begin{figure}
 \centering
 \includegraphics[width=\columnwidth]{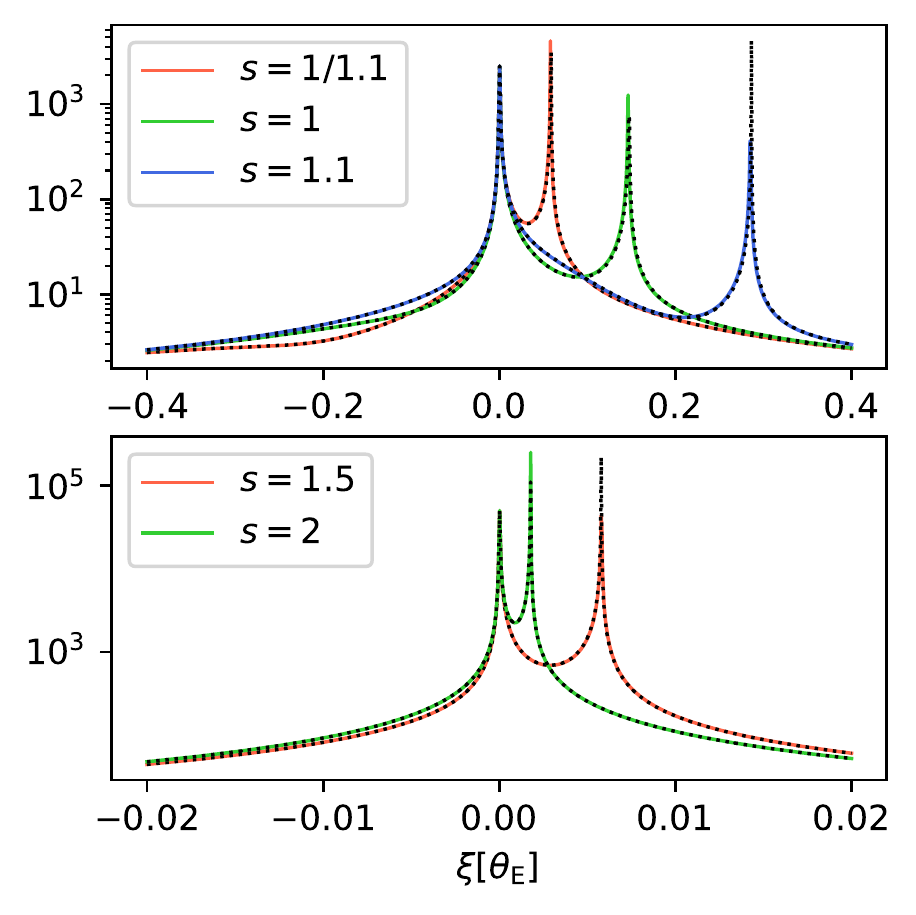}
 \caption[Real-axis magnifications under the variable-shear and exact calculations]{Real-axis magnifications under the variable-shear (solid lines) and exact (dotted lines) calculations. The top panel corresponds to the lens configurations in Figure \ref{fig:mag-resonant} and the bottom panel corresponds to the configurations in Figure \ref{fig:mag-central}. The vertical axes show magnification on log scale.}
 \label{fig:mag-axis}
\end{figure}

In contrast, the right two panels of Figure \ref{fig:lc-central} show that variable-shear magnifications are much more accurate in this ultra high-magnification regime for major-image perturbations, including inside of central and resonant caustics (also see Figure \ref{fig:illus1}a).
Lastly, despite these anomalous behaviors in the regime of $u_0\sim q$, Figure \ref{fig:mag-axis} shows that the variable-shear calculation is nearly identical to the exact calculation on the real axis in the high-magnification regime.
This behavior may be straightforwardly derived using the resultant method \citep{witt_minimum_1995}, which was adopted in \cite{zhang_mathematical_2022} to derive closed-form magnifications for the planetary lens on the real axis.

\section{Caustics}
\label{sec:caus}

Let us first examine the interpretation of central and resonant caustics under the variable-shear Chang-Refsdal lens approximation, which will also assist the interpretation of Figure \ref{fig:mag-central}.
Imagine a hypothetical source on the real axis moving across the primary star from $\xi>0$ to $\xi<0$. Here, the shear changes from $\gamma<1$ to $\gamma>1$, and the underlying Chang-Refsdal caustic splits into two (compare Figure \ref{fig:illus1}a/b). 
Since the horizontal size of the $\gamma>1$ Chang-Refsdal caustic diverges to infinity in the limit of $\gamma\rightarrow1$, this hypothetical source transitions from the inside to the outside of caustics in the above scenario, thus resulting in a cusp exactly at the primary star, which is referred to the primary cusp.
Moreover, since the two caustic folds originating from the primary cusp are parts of $\gamma>1$ Chang-Refsdal caustics, the back ends of the central/resonant caustic are restricted to the negative source plane. In comparison, there is always a small offset between the primary cusp and the primary star under the exact calculation, where the primary folds are also allowed to traverse into the positive source plane (see Figure \ref{fig:mag-central}).

For the remainder of this section, I will examine how the variable-shear Chang-Refsdal lens approximation recovers known caustic properties of the two-body planetary lens. Caustic cusp locations can be derived under the variable-shear approximation by recognizing that Chang-Refsdal cusp locations are expressed as simple analytic expressions of the shear (e.g., \citealt{an_changrefsdal_2006}), which are related back to the cusp locations themselves via Equation \ref{eq:shear}.
Here, I will examine two special cases: the central caustic and the $s=1$ resonant caustic.

Since the primary cusp is always located at the primary star and coordinate origin under the variable-shear formalism, the length of the central caustic is given by the location of its other cusp on the real axis located at $\xi_c$. Since $\xi_c\ll1$, the shear at $\xi_c$ is
\begin{equation}
\label{eq:cc}
    \gamma_c=\dfrac{1}{z_+^2(\xi_c)}\simeq 1-\xi_c,
\end{equation}
which is illustrated in the variable-shear axis labels in Figure \ref{fig:mag-central}.
The real-axis cusps for $\gamma<1$ Chang-Refsdal caustics are located at $\pm2\gamma/\sqrt{1-\gamma}$ in the planetary frame. Equating its locations in the primary and planetary frames (Equation \ref{eq:trans}) and substituting in the shear in Equation \ref{eq:cc}, we arrive at
\begin{equation}
\label{eq:central}
    \sqrt{q}\left(s-\dfrac{1}{s}-\xi_c\right)=\pm \dfrac{2\gamma_c}{\sqrt{1-\gamma_c}}=\pm \dfrac{2(1-\xi_c)}{\sqrt{\xi_c}},
\end{equation}
where the plus sign corresponds to the wide topology and the minus sign for the close topology. The above equation can be rearranged into a cubic polynomial in $\sqrt{\xi_c}$. We may then Taylor expand the valid cubic root in $q$ and acquire
\begin{equation}
\label{eq:xic}
    \xi_c=\dfrac{4q}{(s-1/s)^2}-\dfrac{32q^2s^4(s^2-s-1)}{(s^2-1)^5}+\bigo{q^3}.
\end{equation}

The first order $q$ term is invariant under $s\leftrightarrow1/s$ and is in agreement with the exact planetary lens (e.g., \citealt{an_gravitational_2005,chung_properties_2005}). However, the second-order term disagrees (e.g., \citealt{an_condition_2021}; Eq.\ 13), indicating higher-order differences. Note that by clearing fractions in Equation \ref{eq:central} and dropping the highest order term in $\xi_c$, Equation \ref{eq:central} itself becomes invariant under $s\leftrightarrow1/s$, allowing one to directly acquire the first-order-$q$ term without the Taylor expansion.

For the $s=1$ resonant caustic, the origins for the primary and primary coordinates coincide ($s-1/s=0$), indicating that the ``planetary cusps'' are exactly on the imaginary axis of the primary frame, where the shear becomes $\gamma=1$. Therefore, the imaginary-axis cusps are located at $\eta^{[2]}=\pm2\gamma/\sqrt{1+\gamma}=\pm\sqrt{2}$ in the planetary frame, and $\eta_r=\pm\sqrt{2q}$ in the primary frame. The vertical size of the $s=1$ caustic is therefore $\Delta\eta_r=2\sqrt{2q}$.

The horizontal size of the $s=1$ caustic may be derived in a similar manner as the central caustic via Equation \ref{eq:central}. With the location of the real-axis resonant-caustic cusp written as $\xi_r$ with shear $\gamma_r$,
\begin{equation}
    \sqrt{q}\cdot\xi_r=\dfrac{2\gamma_r}{\sqrt{1-\gamma_r}}.
\end{equation}
Substituting in Equation \ref{eq:shear} and expanding up to first order in $\xi_r$, we have
\begin{equation}
    \dfrac{2}{\sqrt{\xi_r}}-\dfrac{3\sqrt{\xi_r}}{2}-\dfrac{\xi_r}{\sqrt{q}}=0.
\end{equation}
For $\xi_c\ll1$ and $q\ll1$, the $\sqrt{\xi_c}$ term may be dropped, which result in $\xi_r=\sqrt[3]{4q}$, and this is the length of the resonant caustic. The above results also show that the vertical-to-horizontal width ratio of the resonant caustic scales as
\begin{equation}
    \eta_r/\xi_r\propto q^{1/6},
\end{equation}
which does not appear to be well-known in the literature.

\section{Semi-Analytic Solutions}
\label{sec:semi}
In this section, I demonstrate how the generalized perturbative picture allows the full two-body lens equation to be solved semi-analytically. Given that the image in the negative lens plane is only weakly affected by the planet, its unperturbed location
\begin{equation}
\label{eq:init}
    z_{\rm PSPL}=\dfrac{\zeta}{2}\cdot\left(1\pm\sqrt{1+4|\zeta|^{-2}}\right)
\end{equation}
can be used as an initial guess to Newton's or Laguerre's method to quickly solve for one quintic root of the lens equation. Here, PSPL refers to point-source point-lens. In the above equation, the plus sign represents the major image location that is chosen for minor image perturbations, and vice versa.
Once one quintic root is found and divided out, the resulting quartic polynomial can be solved in closed form. The quartic roots can then be verified with the full quintic equation and, depending on the requested precision, may be optionally refined by Newton's method to reduce the numerical noise from the initial root division. This noise is nevertheless expected to be small for well-isolated roots (e.g.\ \citealt{skowron_general_2012}). Indeed, the weakly perturbed image is also generally the most isolated image (Figure \ref{fig:perturb}), and thus the final refinement may not be necessary.

Note that the closed-form quartic solution discovered by Lodovico Ferrari is known to suffer from certain round-off errors for cases with large root spread (e.g.\ \citealt{strobach_fast_2010}), defined as the ratio between the largest and smallest root magnitudes. Therefore, the coordinate origin is defined at the primary star, as only the minor image becomes close to the origin for very faraway sources. In comparison, other frameworks such as VBBL \citep{bozza_vbbinarylensing_2018} have also considered coordinate origins at the planetary location, which may induce large root spread as one image is usually very close to the planet.
In future work, an improved quartic solver proposed by \cite{orellana_algorithm_2020} may also be explored, which is robust against these errors but only costs twice the computational time as Ferrari's solution.

The initial root-refinement step may benefit from a combination of Newton's and Laguerre's methods depending on the polynomial residual \citep{skowron_general_2012}, which is larger for sources near the imaginary axis in our case.
As an illustration, in the case of the $s=1$ magnification map in Figure \ref{fig:mag-resonant}, it only took 2/4 iterations with Laguerre's/Newton's method to refine from the PSPL location of the weakly affect image location to a polynomial residual less than $10^{-14}$ for $99.9\%$ of the pixels.
In comparison, using the PSPL location of the strongly perturbed image, the planet location, and the primary location takes 7/16, 5/16, and 9/23 iterations\footnote{These numbers also suggest an alternative and potentially useful approach to first find and divide out three roots initializing from the two PSPL locations and the planet location, where the resulting quadratic equation after root division could then be solved in closed form.} with Laguerre's/Newton's method to locate the first quintic root subject to the same precision requirements, which demonstrates the comparative advantage of starting from the weakly affected image.

A basic benchmark test of a vectorized python implementation provided in the code repository of this paper shows that $s=1$ magnification map in Figure \ref{fig:mag-resonant} with $2.5\times10^5$ pixels is calculated in merely 0.6s with the semi-analytic method, with the two steps of finding the initial root and solving the quartic polynomial in closed-form taking around 0.3s each. Since the total cost is only twice the cost of a few iterations of Newton's method to solve for the initial root, we may expect the semi-analytic method to be substantially faster than the standard numerical approach (e.g.\ \citealt{skowron_general_2012}). However, we do not attempt to quantify the factor of speed-up here, which involves the delicate task of holding the level of optimization consistent across all methods tested.
In comparison, the analytic variable-shear solution of Section \ref{sec:mag} costs only 0.2s for the same $s=1$ magnification map. The semi-analytic solver also applies to binary mass ratios, which takes a slightly longer 1s for $q=0.9$ and $s=1$ given the reduced accuracy of Equation \ref{eq:init} as the initial guess. The above numbers will be dependent on the computational device and may be rerun with the code-base provided.

\section{Conclusions}
\label{sec:disc}
In this paper, I have introduced the idea of a generalized perturbative picture for planetary microlensing, which states that the planet acts as a variable-shear Chang-Refsdal lens on one of the unperturbed images in the positive lens plane, leaving the other image largely unaffected.
The proposed framework generalizes upon the original perturbative picture of \cite{gould_discovering_1992} and the uniform-shear Chang-Refsdal lens approximation of \cite{gaudi_planet_1997}, by relaxing both the required proximity between the planet and the image being perturbed and the condition of isolated planetary caustics.

Under the generalized perturbative picture, the action of the planet can be classified into major image perturbations and minor image perturbations, which are distinguished by whether the source is in the positive ($\xi>0$) or negative ($\xi<0$) source plane, as opposed to the location of the planet being inside ($s<1$) or outside ($s>1$) the Einstein ring of the primary star (cf.\ Appendix in \citealt{han_moa-2016-blg-319lb_2018}).
Moreover, the generalized perturbative picture demonstrates that the existence of a unified regime of light-curve degeneracy independent of caustic topologies can be explained by the symmetry of the Chang-Refsdal lens. Therefore, the offset degeneracy can be interpreted as a generalization of the inner-outer degeneracy for planetary caustics, both of which describe an ambiguity as to ``whether the planet lies closer to or farther from the star than does the position of the image that it is perturbing'' \citep{gaudi_planet_1997}.

It should be noted that the variable-shear Chang-Refsdal lens approximation is proposed without formal derivation in this paper. Instead, I have demonstrated that the proposed formalism not only produces accurate magnification maps (Section \ref{sec:mag}), but also recovers known caustic properties of the full planetary lens (Section \ref{sec:caus}). An interesting property of the variable-shear lens is that it does not recover the correct image positions, despite its accuracy in the source plane.
This intriguing behavior deserves further analytical study in future works.

Moving forward, it is beneficial for the two analytic prescriptions (Section \ref{sec:mag} \& \ref{sec:semi}) together with finite-source algorithms (e.g.\ \citealt{dominik_robust_1998,gould_hexadecapole_2008,bozza_microlensing_2010}) to be implemented\footnote{I note that one such differentiable microlensing code named \texttt{caustics} (Bartoli\'c \& Dominik, in prep), is currently under development.} in automatic-differentiation frameworks such as \texttt{jax} \citep{bradbury_jax_2018} or \texttt{julia} \citep{bezanson_julia_2017}, which allows the gradient of the likelihood function to be acquired without deriving explicit expressions.
This allows for the use of gradient-based inference algorithms, particularly Hamiltonian Monte Carlo (HMC) methods including the No-U-Turn Sampler (NUTS; \citealt{hoffman_no-u-turn_2014}), which utilize gradient information to avoid the random walking behavior of common Markov chain Monte Carlo (MCMC) samplers such as Metropolis-Hastings. For the exact semi-analytic approach, it is also not necessary for the gradient to be ``back-propagated'' through the root-refinement step for planetary mass ratios, where the location of the weakly affect image is insensitive to the planetary parameters.

\section*{Acknowledgements}
I would like to thank Scott Gaudi and Jin An for helpful discussions, and Joshua Bloom, Scott Gaudi, Jessica Lu, Sean Terry, and Weicheng Zang for comments on the manuscript.
K.Z.\ thanks the LSSTC Data Science Fellowship Program, which is funded by LSSTC, NSF Cybertraining Grant \#1829740, the Brinson Foundation, and the Moore Foundation; his participation in the program has benefited this work.
This work is partially supported by funding from a Two Sigma faculty fellowship and the Gordon and Betty Moore Foundation.

\section*{Data availability}
The calculations made in this paper can be reproduced using code provided at https://github.com/kmzzhang/analytic-lensing.
\bibliographystyle{mnras}

\bsp	
\label{lastpage}
\end{document}